# Implications of Lorentz covariance for the guidance equation in two-slit quantum interference

**Peter Holland**
Green College
University of Oxford
Woodstock Road
Oxford OX2 6HG
England

peter.holland@green.ox.ac.uk

**Chris Philippidis**
Science Studies Centre
University of Bath
Bath BA2 7AY
England

c.philippidis@bath.ac.uk

**Abstract**

It is known that Lorentz covariance fixes uniquely the current and the associated guidance law in the trajectory interpretation of quantum mechanics for spin $\frac{1}{2}$ particles. In the non-relativistic domain this implies a guidance law for the electron which differs by an additional spin-dependent term from that originally proposed by de Broglie and Bohm. In this paper we explore some of the implications of the modified guidance law. We bring out a property of mutual dependence in the particle coordinates that arises in product states, and show that the quantum potential has scalar and vector components which implies the particle is subject to a Lorentz-like force. The conditions for the classical limit and the limit of negligible spin are given, and the empirical sufficiency of the model is demonstrated. We then present a series of calculations of the trajectories based on two-dimensional Gaussian wave packets which illustrate how the additional spin-dependent term plays a significant role in structuring both the individual trajectories and the ensemble. The single packet corresponds to quantum inertial motion. The distinct features encountered when the wavefunction is a product or a superposition are explored, and the trajectories that model the two-slit experiment are given. The latter paths exhibit several new characteristics compared with the original de Broglie-Bohm ones, such as crossing of the axis of symmetry.

# 1 Introduction

The interpretation of quantum mechanics pioneered by de Broglie [1] and Bohm [2] showed how a deterministic picture of particle trajectories can consistently underpin the statistical predictions of quantum mechanics. In the non-relativistic case de Broglie and Bohm based their interpretation on the particle law of motion (the 'guidance' equation)

$$m\dot{\mathbf{x}} = \nabla S \tag{1.1}$$

where $S$ is the phase of the Schrödinger wavefunction ($\Psi = \sqrt{\rho} \exp(iS/\hbar)$). Interest in this field was stimulated by plots of the trajectories $\mathbf{x} = \mathbf{x}(t, \mathbf{x_0})$ implied by (1.1) for electrons passing through a two-slit interferometer [3]. Over an ensemble, the characteristic interference pattern is built up through channelling of the paths into bands due to action by the quantum potential (Fig. 1). The patterns obtained remain among the most striking illustrations so far of the insight provided by this theory into quantum phenomena. The original computations provided a paradigmatic example for the calculation as well as the reading of the constructs of the de Broglie-Bohm theory.

Recently, however, it has become apparent that there is a problem of underdetermination in the de Broglie-Bohm theory [4,5]. The model works in the first instance because the flow implied by the law (1.1) generates the quantal distribution $\rho = |\Psi|^2$ at time $t$ if the initial distribution is given by the quantal expression $\rho_0 = |\Psi_0|^2$. This follows immediately from the conservation equation implied by the Schrödinger equation:

$$\frac{\partial \rho}{\partial t} + \nabla \cdot \mathbf{j}_S = 0, \quad \mathbf{j}_S = \rho \nabla S/m \tag{1.2}$$

The difficulty is that there is a "gauge" freedom in the current: any vector $\mathbf{j}$ that differs from the traditional choice $\mathbf{j}_S$ by a divergenceless vector field $\mathbf{a}$,

$$\mathbf{j} = \mathbf{j}_S + \mathbf{a}, \quad \nabla \cdot \mathbf{a} = 0, \tag{1.3}$$

will define a current compatible with the same density of particles. Hence, one may choose for the guidance law the expression

$$\dot{\mathbf{x}} = \mathbf{j}/\rho \tag{1.4}$$

and still maintain empirical compatibility with quantum mechanics. The basic requirement that, for an ensemble of particles, the flow reproduces exactly the quantal position probability distribution is not sufficiently restrictive to fix $\mathbf{a}$. The problem is not solved by appealing to Hamilton-Jacobi theory, as was employed by Bohm - indeed, formulating the de Broglie-Bohm theory in Hamiltonian terms suggests yet further classes of alternative guidance equations [6]. Some further natural restriction(s) on the possibilities for defining the particle law of motion is required.

One may envisage a variety of physically reasonable constraints which may potentially limit the possibilities for $\mathbf{a}$. One set of constraints, pertaining to Galilean invariance, nonlocality, and empty waves, has been shown not to be sufficiently restrictive to yield a unique expression [5]. This set is not, however, exhaustive and one can advance other arguments, such as the requirement that in interactions with an electromagnetic field it is the usual Schrödinger current that acts as a source for the field, which imply that $\mathbf{a} = 0$. Hence, one can invoke arguments that uniquely fix $\mathbf{j}$, and indeed in the case just cited this is just $\mathbf{j}_S$ (so that the de Broglie-Bohm law (1.1) will then be unique).

In this paper we adopt an approach which offers an answer to this problem without invoking the details of interactions with other fields. In this connection, we note that conceptually the trajectories of the original de Broglie-Bohm model are not just a network of extremals (a property they inherit from the Hamilton-Jacobi equation [6]). They also have the property of connecting the points of the manifold of $\rho$ convectively (the continuity equation gives them the character of a congruence of streamlines), and they claim their ontological status from both these properties. The original trajectories were computed from the gradient of the phase alone and emphasised the particle aspect at the expense of that of the field. Although both the fields $S$ and $\rho$ are implicated in the dynamics (since they are coupled by the Schrödinger equation), $S$ alone appears explicitly in the guidance equation. The dynamical role of $\rho$ was represented indirectly through the quantum potential:

$$Q(\mathbf{x},t) = -\frac{\hbar^2}{2m\sqrt{\rho}}\nabla^2\sqrt{\rho} = \frac{\hbar^2}{4m\rho}\left(\frac{1}{2\rho}(\nabla\rho)^2 - \nabla^2\rho\right). \tag{1.5}$$

Bohm gave $Q$ an informational interpretation, and it is arguable that this was necessary because of the segregation of $S$ and $\rho$. It seems reasonable to enquire whether a formulation can be found that puts $S$ and $\rho$ on a more equitable dynamical footing.

The approach we shall take to the problem of fixing the non-relativistic guidance equation is one that naturally implies a common dynamical role for both components of the wavefunction. In this approach, the non-relativistic uniqueness problem is solved by first solving the analogous problem in relativistic quantum theory. It has been shown recently by one of us [7] that, for spin $\frac{1}{2}$ particles, the requirements imposed by Lorentz covariance are sufficient to determine the relativistic particle law of motion uniquely. The assumption that the non-relativistic guidance equation is a limit of the relativistic one then uniquely fixes that law, too. The equation of motion obtained in the latter case differs from (1.1) by a spin-dependent addition on the right-hand side. This additional term brings the density explicitly into the guidance law.

The aim of the present paper is to explore some of the implications of the revised non-relativistic guidance equation, and to apply it to the iconic case study of two-slit interference with a view to re-examining the calculations and readings of the de Broglie-Bohm model. The new guidance equation is derived in §2, and in §§3-5 some of its general implications are investigated, pertaining to mutual dependence of orthogonal motions, the role of the quantum potential and force, and limiting cases. The empirical sufficiency of the model is demonstrated in §6. We then present a series of calculations of the paths that illustrate some of the key novel features of the revised guidance law. Specifically, in §§8 and 9 we examine a two-dimensional Gaussian wave packet and find that the additional spin-dependent term plays a significant role in structuring both the individual trajectories and the ensemble. In particular, this wavefunction implies inertial motion. The distinct features encountered when the wavefunction is a product or a superposition are explored in §10, and the trajectories that model the two-slit experiment are presented in §11. The latter paths exhibit several new characteristics compared with the original ones (Fig. 1), an example being that they cross the axis of symmetry.

## 2 Uniqueness of the non-relativistic guidance equation

The argument for uniqueness follows from the combination of Lorentz covariance and the probability interpretation of the wavefunction (in a suitably low-energy regime where this is meaningful) [7]. For a spin $\frac{1}{2}$ particle whose associated spinor field $\psi^a(x^i,t)$ obeys the Dirac equation, the Dirac current satisfies the conservation law

$$\partial_\mu j^\mu = 0, \quad j^\mu = c\bar{\psi}\gamma^\mu\psi, \quad \bar{\psi} = \psi^+\gamma^0. \tag{2.1}$$

To establish the uniqueness of the Dirac current, consider a new current $\bar{j}^\mu$ which differs from $j^\mu$ by the addition of a divergenceless 4-vector $a^\mu$:

$$\bar{j}^\mu = j^\mu + a^\mu, \quad \partial_\mu \bar{j}^\mu = 0. \tag{2.2}$$

The new current is conserved and so could potentially be used to define a probability 4-vector. In order that the new current still reproduces the quantal distribution, we must have $a^0 = 0$. Now consider the current in another Lorentz frame, with spinor $\psi'(x'^i, t')$. Then in this frame $j'^0 = c\psi'^+\psi'$ and again $a'^0 = 0$. But the only 4-vector whose zeroth component vanishes in all frames is the zero vector. Hence $a^\mu = 0$ and the Dirac current is unique. This demonstration can be extended to the many-particle case [7]. (It is assumed in this proof that the correct expression for the zeroth component of the relativistic current has already been identified. An examination of the possibility of an *ab initio* proof of the uniqueness of the Dirac current that proceeds without first making this assumption is given elsewhere [8]. There it is shown that the Dirac current is the only conserved 4-vector implied by the Dirac equation that is a function of just the quantum state.)

In the non-relativistic limit the Dirac equation passes over to the Pauli equation for a 2-component spinor $\phi^A$, A = 1,2, and the limiting Pauli current implied by the Dirac current is

$$\rho\left(\equiv j^0/c\right) = \phi^+\phi \tag{2.3}$$

$$\mathbf{j} = (\hbar/2mi)\left(\phi^+\nabla\phi - (\nabla\phi^+)\phi\right) - (e\rho/mc)\mathbf{V} + (1/m)\nabla\times(\rho\mathbf{s}) \tag{2.4}$$

where $\mathbf{s} = (\hbar/2\rho)\phi^+\boldsymbol{\sigma}\phi$ is the spin vector and $\mathbf{V}$ is the external vector potential. The current (2.4) comprises convection (the first two terms) and spin contributions. It is "non-relativistic" in that it transforms in the expected way with respect to Galilean boosts. That is, under the transformation $\mathbf{x}' = \mathbf{x} - \mathbf{u}t$, $t' = t$ we have $\rho'(\mathbf{x}',t') = \rho(\mathbf{x},t)$, $\mathbf{j}'(\mathbf{x}',t') = \mathbf{j}(\mathbf{x},t) - \mathbf{u}\rho(\mathbf{x},t)$.

For the special case where the magnetic field may be neglected and the system is in a spin eigenstate,

$$\phi^A(\mathbf{x},t) = \Psi(\mathbf{x},t)\chi^A, \quad \chi^+\chi = 1, \tag{2.5}$$

the Pauli equation reduces to the Schrödinger equation for the function $\Psi$ but the current (2.4) does not coincide with the conventional Schrödinger expression for the current, $\mathbf{j}_S$ in (1.2) [9,10]. Writing $\Psi = \sqrt{\rho}\exp(iS/\hbar)$ we have

$$\mathbf{j} = (1/m)\rho\nabla S + (1/m)\nabla\rho\times\mathbf{s}, \quad \mathbf{s} = (\hbar/2)\chi^+\boldsymbol{\sigma}\chi, \tag{2.6}$$

which contains a generally non-vanishing, spin-dependent term.

It follows from these considerations that the formulas determining the trajectory of a spin $\frac{1}{2}$ particle in the relativistic and non-relativistic cases are unique (if the latter is considered as a limiting case of the former). In the former case the guidance equation is given by [11,12]

$$\frac{dx^i}{dt} = \frac{c\, j^i}{j^0}, \quad i = 1, 2, 3. \tag{2.7}$$

Varying $x_0^i = x^i(t=0)$, the flow $x^i(t)$ reproduces the distribution $j^0 = c\psi^+\psi$ given an initial distribution $\psi_0^+\psi_0$ (note that this law is consistent in that the Dirac current is timelike with a positive zeroth component). In the non-relativistic regime, the expression (2.4) gives for the guidance equation (neglecting the magnetic field)

$$m\dot{\mathbf{x}} = (\hbar/2i\rho)(\phi^+\nabla\phi - (\nabla\phi^+)\phi) + (1/\rho)\nabla\times(\rho\mathbf{s}). \tag{2.8}$$

Finally, for a spin eigenstate, we obtain from (2.6)

$$m\dot{\mathbf{x}} = \nabla S + \nabla\log\rho\times\mathbf{s}. \tag{2.9}$$

Since it is the limiting case of the (unique) Dirac current, there is no freedom to add a divergence-free vector field to (2.8) (or (2.9)), that is, this non-relativistic law for spin ½ particles is unique[1]. In connection with an analogy with charged particle motion in an electromagnetic field we shall explore later (§4), it is convenient to write the spin-dependent addition in (2.9) as a "vector potential":

$$m\mathbf{v} = \nabla S - \mathbf{A}, \quad \mathbf{A} = -\nabla\log\rho\times\mathbf{s}. \tag{2.10}$$

It follows from $\nabla.(\rho\mathbf{A}) = 0$ and $\nabla\log\rho.\mathbf{A} = 0$ that $\nabla.\mathbf{A} = 0$, so $\mathbf{A}$ is a solenoidal vector.

Comparing with the de Broglie-Bohm version of the non-relativistic law of motion, (1.1), we see that both laws are compatible with the same density of particles $\rho$ (for the spin contribution to the current (2.4) has vanishing divergence) but (1.1) is inconsistent if the theory is to be considered as embedded ultimately in a relativistic spin ½ theory. Rather, the domain of validity of the de Broglie-Bohm theory is restricted to spin 0 particles. We emphasize that the spin-dependent addition must be present even in the absence of magnetic fields. It is the guidance law (2.9), which holds for a spin eigenstate, that is the object of enquiry in this paper. Note that, as is the case with the de Broglie-Bohm guidance equation, we need only specify the initial position of the particle to compute trajectories from (2.9).

Although there exist arguments for introducing the spin current within non-relativistic quantum mechanics [13], it does not seem that one can prove the uniqueness of the total current so obtained in that context (i.e., one can potentially add still further divergenceless currents without disturbing the probability density). In view of the fact that, as demonstrated above, relativity requires that this term (and only this term) be present in addition to the usual convection current, it seems reasonable to regard the existence and properties of the spin-dependent deviation from (1.1) as residues of the relativistic theory in the non-relativistic limit.

The presence of the spin term is significant also in that it gives a concrete example of what happens to the trajectories when a divergenceless "gauge" vector is added to the Schrödinger current. However, unlike the class of gauges envisaged by, for example, Deotto and Ghirardi [5], the additional term here is uniquely selected from all possible ones by a physical constraint, namely

---

[1] It should be noted that the form of uniqueness demonstrated here refers to flows that directly generate the particle probability density in the position representation. This does not exclude the possibility that $m\mathbf{j}/\rho$ may emerge as, for example, the mean momentum in some other theory that introduces additional variables to describe the motion. For purposes of comparison we recall that in classical statistical mechanics the analogue of equation (1.2) is not basic but follows as a space projection of the phase space Liouville equation.

Lorentz covariance. That is, this physical principle provides a reason for a particular choice of the "gauge" and at the same time removes the freedom to add any other terms to the velocity.

The role played by the spin term in the context of trajectory interpretations of quantum mechanics has been analysed previously [9,10,12]. Its role in the definition of arrival times has been considered [14], and the orbits for Hydrogen atom eigenstates have been discussed [15]. In addition, Dirac flow lines (which of course tacitly exhibit the effect of this term) have been plotted [11,16,17]. To our knowledge, the theory has not hitherto been applied to an examination of the effect of the spin term on quantum trajectories in the case of Gaussian packets and the double-slit.

## 3 Kinetic interdependence in states with product wavefunctions

A striking property of the spin-dependent addition is that the components of the particle motion in orthogonal directions are generally mutually dependent, even when the wavefunction factorizes in these directions. Suppose for simplicity that $\mathbf{s} = (\hbar/2)(0,0,1)$ (this is always assumed in our calculations in §§8-11). Then (2.9) becomes

$$m\dot{x} = \frac{\partial S}{\partial x} + \frac{\hbar}{2\rho}\frac{\partial \rho}{\partial y}, \quad m\dot{y} = \frac{\partial S}{\partial y} - \frac{\hbar}{2\rho}\frac{\partial \rho}{\partial x}, \quad m\dot{z} = \frac{\partial S}{\partial z}. \qquad (3.1)$$

When the wavefunction factorizes, $\Psi = \Psi_1(x)\Psi_2(y)\Psi_3(z)$, we obtain

$$m\dot{x} = \frac{\partial S_1}{\partial x} + \frac{\hbar}{2\rho_2}\frac{\partial \rho_2}{\partial y}, \quad m\dot{y} = \frac{\partial S_2}{\partial y} - \frac{\hbar}{2\rho_1}\frac{\partial \rho_1}{\partial x}, \quad m\dot{z} = \frac{\partial S_3}{\partial z} \qquad (3.2)$$

from which our assertion is obvious (clearly, this is not an artefact of the particular choice made here for **s**). This property of kinetic interdependence is analogous to but not the same as quantum-mechanical entanglement. It is similar to a constraint that couples orbital degrees of freedom that are *statistically* independent, as implied by a factorized wavefunction. Rather like the vanishing of the internal constraints of a rigid body on a macroscopic average, the kinetic interdependence does not in general show up explicitly in ensemble averages.

The equations (3.2) illustrate that we have to be careful if we desire to extract the description of a purely one-dimensional system (say, *z*) from the full three-dimensional theory. The usual condition for this, that the wavefunction factorizes, is sufficient only if the spin vector lies along the relevant direction (i.e. *z*). Then the guidance law for *z(t)* is independent of the *x* and *y* coordinates, and indeed is the usual de Broglie-Bohm equation (as it must be since, as stated above, the de Broglie-Bohm law is the unique guidance equation in one dimension).

A further consequence of (3.2) has to do with parametrization of the temporal evolution of an orbit by a space coordinate. Suppose the wavefunction is a packet in the *y*-direction and a plane wave in the *x*-direction. Then according to the de Broglie-Bohm formula (1.1) *x(t)* is proportional to *t*. Hence, instead of *t,* we may parametrize the orbits *y(t)* by *x* and obtain trivially the corresponding orbits in the *xy*-plane, apart from a scaling factor (this was done in the original two-slit calculations [3]). In contrast, using the full formulas (3.2) to describe motion in the *xy*-plane we clearly cannot parametrize the paths in this way since the *x*-coordinate will generally be a complicated function of *t*. Although for this quantum state the orbits *y(t)* will be the same (as the de Broglie-Bohm ones), the space curves *y(x)* will be very different.

A final implication of (3.2) has to do with the assertion of Hestenes [9,10] and Bohm and Hiley [12] that "spin" should not be attributed to the particle but is rather to be regarded as due to an additional circulatory feature of the orbit implied by the spin term in the current. In this paper we are concerned purely with the translational motion of the particle and shall not address the issue of how spin angular momentum is to be interpreted (i.e., whether spin is a property of the particle, or of the guiding wave acting on a point particle). We do, however, remark that the effect of the spin current on the orbit is not in fact generally circulatory. This will be so even if $\rho_1$ and $\rho_2$ in (3.2) have the same functional forms (see §8).

Insight into the meaning of spin may be gained by examining its role in relation to the ensemble rather than to an individual orbit. To this end, from (2.9) we define the total particle orbital angular momentum to be $\mathbf{L} = \mathbf{x} \times \mathbf{p}$ where $\mathbf{p} = \nabla S + \nabla \log \rho \times \mathbf{s}$. It is thus composed of two parts:

$$\mathbf{L} = \mathbf{L}_{conv} + \mathbf{L}_{spin}, \quad \mathbf{L}_{conv} = \mathbf{x} \times \nabla S, \quad \mathbf{L}_{spin} = \mathbf{x} \times (\nabla \log \rho \times \mathbf{s}). \tag{3.3}$$

By partial integration, one can show that the mean angular momentum coming from the spin contribution is simply (2x) the spin vector:

$$\langle \mathbf{L}_{spin} \rangle = \int \rho \mathbf{L}_{spin} d^3 x = 2\mathbf{s}. \tag{3.4}$$

We expect therefore from (3.4), which is true for all wavefunctions, that an ensemble of trajectories will exhibit an overall circulatory behaviour due to the collective effect of the vector potential. This feature is indeed illustrated in our examples.

## 4 The new quantum potentials and the force law

Part of the attraction of the quantum potential (1.5) as an explanatory device in the original de Broglie-Bohm model (based on the guidance formula (1.1)) is the simplicity of its form, and that it encodes in a single function the difference between classical and quantum behaviour (although this is a subtle question; see [18]). This follows from the quantum Hamilton-Jacobi equation,

$$\frac{\partial S}{\partial t} + \frac{1}{2m}(\nabla S)^2 + Q + V = 0 \tag{4.1}$$

(the sense in which this may be regarded as a genuine Hamilton-Jacobi equation is discussed in [6]), and the generalization of Newton's second law that is implied by (4.1) on taking the spatial gradient:

$$m\ddot{\mathbf{x}} = -\nabla(Q + V). \tag{4.2}$$

The question arises whether one can locate the analogue(s) of the quantum potential in the context of the revised guidance formula (2.9), and whether the concept of a quantum potential remains useful. To this end, we rewrite (4.1) (which remains valid in the present case) in the following form:

$$\frac{\partial S}{\partial t} + \frac{1}{2m}(\nabla S - \mathbf{A})^2 + Q' + V = 0, \quad \mathbf{A} = -\nabla \log \rho \times \mathbf{s}. \tag{4.3}$$

Here we have introduced the "vector potential" expression (2.10) for the momentum in the kinetic energy term, and written

$$Q' = Q + \frac{1}{m}\nabla S \cdot \mathbf{A} - \frac{1}{2m}\mathbf{A}^2. \tag{4.4}$$

Clearly, the latter term plays an analogous role to the original quantum potential. However, unlike the de Broglie-Bohm model the scalar $Q'$ does not fully represent the quantum effects on the particle, as these are now also to be attributed to **A** (for we recover the classical Hamilton-Jacobi equation only when both $Q'$ and **A** are negligible). We therefore must consider both scalar and vector quantum potentials. It will be seen that **A** carries information on the directional behaviour of the first derivatives of the field $\rho$, while $Q'$ encodes the finer details of the spatial variation, given by the second derivatives.

To obtain a useful expression for the effects of these potentials, we derive the generalization of Newton's law implied by (4.3). This follows immediately by analogy with the derivation of the Lorentz force law from the Hamilton-Jacobi equation for a classical charged particle in electrodynamics:

$$m\ddot{\mathbf{x}} = \mathbf{E} + \dot{\mathbf{x}} \times \mathbf{B} - \nabla V \tag{4.5}$$

with

$$\mathbf{E} = -\nabla Q' - \frac{\partial \mathbf{A}}{\partial t}, \quad \mathbf{B} = \nabla \times \mathbf{A} \tag{4.6}$$

where on the right-hand side of (4.5) we insert $\mathbf{x} = \mathbf{x}(t)$ in the fields. We see then that, apart from the classical force, the particle is subject to a quantum force that is identical in form to the Lorentz force in classical electrodynamics, with the quantum potentials playing the role of the electromagnetic potentials. This differs significantly from the Bohm force law (4.2) (to which it would reduce were **A** = 0) but it seems that the notions of "quantum potential" and the corresponding quantum force remain useful. It will be convenient to refer to **E** and **B** as "electric-like" and "magnetic-like" fields, respectively[2].

An analysis based on the relation between the property of kinetic interdependence described in §3 and the Lorentz-like force has some significant consequences. First, as we shall see in §8.1, it can cause the Lorentz-like force to vanish. Second, for the case of factorized wavefunctions with guidance formula (3.2), the only component of the Lorentz force acting in the $z$-direction comes from the component $-\nabla Q$ of **E**, corresponding to the usual de Broglie-Bohm quantum force (4.2). For example, by simple vector algebra we can write the magnetic field in general as follows

$$\mathbf{B} = -\nabla(\nabla \log \rho \cdot \mathbf{s}) + \mathbf{s}\nabla^2 \log \rho. \tag{4.7}$$

In the case of (3.2) we see that **B** lies along **s** and hence $\dot{\mathbf{x}} \times \mathbf{B}$ lies in the *xy*-plane.

**5 Limit of negligible spin and the classical limit**

We have seen that in the non-relativistic limit of the Dirac theory the special case of a spin eigenstate does not lead to the pure Schrödinger theory; the spin continues to play a key role in the current. There are two cases where the effect of spin will be negligible in the current. First, the additional term will vanish exactly when $\nabla \rho$ is proportional to **s**, that is, when the density varies only along the direction defined by **s**. Second, if this condition is not obeyed, we can still obtain the pure Schrödinger current if the wavefunction obeys the following constraint:

$$|\nabla S| \gg \frac{\hbar}{2}|\nabla \log \rho| = \hbar \left|\frac{\nabla \sqrt{\rho}}{\sqrt{\rho}}\right|. \tag{5.1}$$

---

[2] A further interesting question that we shall discuss elsewhere is what quantity plays the role of the electric charge in this analogy. The fields (4.6) have been defined so that (4.5) has a form corresponding to "unit charge", but they each naturally contain the factor $\hbar/2$.

For Schrödinger wavefunctions that satisfy either of these two conditions in a spacetime domain, the guidance equation reduces to the spin 0 one of de Broglie and Bohm[3]. Note though that we do not obtain as a limiting case the full de Broglie-Bohm theory, for the condition (5.1) implies that part of the quantum potential is negligible (see below).

It is interesting to note that the criterion (5.1) was proposed by Dirac [19] as the condition for recovering classical mechanics from quantum mechanics (for spin 0 systems). In fact, this criterion cannot play that role as it is evident from the quantum Hamilton-Jacobi equation (4.1), which holds also for a spin 0 particle, that (5.1) is not sufficient to imply a negligible quantum potential, and hence the classical Hamilton-Jacobi equation. Using the second expression for the quantum potential in (1.5), the criterion (5.1) implies that only part of the quantum potential may be neglected and (4.1) reduces to

$$\frac{\partial S}{\partial t} + \frac{1}{2m}(\nabla S)^2 - \frac{\hbar^2}{4m\rho}\nabla^2 \rho + V = 0 \tag{5.2}$$

and not to the classical Hamilton-Jacobi equation. As we have indicated, (5.1) should rather be interpreted as the condition under which the (dynamical) behaviour of a spin $\frac{1}{2}$ particle approaches that of a spin 0 particle. Since (5.2) contains only a fragment of the quantum potential, we see that the effective spin 0 behaviour obtained differs from that implied in the usual spin 0 theory (described by (4.1)).

Although (5.1) does not characterize the classical limit of a spin 0 system, it is relevant to the classical limit of the spin $\frac{1}{2}$ particle. To arrive at the latter, we require in the first place that the usual quantum potential be negligible [20] so that (4.1) reduces to the classical Hamilton-Jacobi equation:

$$Q \to 0 \tag{5.3}$$

(the notation implies that the quantum force in (4.2) is negligible; the numerical smallness of $Q$ is not sufficient for this). But this will not be sufficient, for the guidance equation (2.9) will still contain quantum effects through the spin-orbit coupling term (which depends on $\hbar$). We must therefore impose in addition the (generally independent) condition (5.1) on the vector quantum potential to achieve classical translational motion. As in (5.3), we require that the derivatives of the vector potential are such that the electric-like and magnetic-like components of the Lorentz force (4.5), **E** and $\dot{\mathbf{x}} \times \mathbf{B}$, are separately negligible relative to the classical force (note that it is not sufficient for classical behaviour that just the total Lorentz force is negligible – see §8.1). We see then that, as regards its translational behaviour, the classical analogue of our quantum system is an ensemble of particles moving in the potential $V$. This limiting procedure will be illustrated later (§§8, 10).

It is emphasized that the limiting behaviours described here refer only to spin 0-like or classical-like behaviour. If, for example, a spin measurement were performed on the particle in either of the limiting states, the wavefunction would become a superposition of spin states (and the spin vector point-dependent) and the tacit quantum nature of the system would become manifest in the particle motion.

---

[3] We are concerned here purely with the translational motion of the particle and do not consider whether the spin $\frac{1}{2}$ and spin 0 theories may be distinguished in other ways (e.g., through the expectation values of operators involving the spin).

## 6 Empirical sufficiency of the model

In order to establish concordance of this model with the measurement results of quantum mechanics we need to demonstrate that the extension of the theory to a system of *N* particles, with wavefunction $\phi^{A_1...A_N}(\mathbf{x}_1,...,\mathbf{x}_N)$, obeys two key properties [20]. The first is that, when the wavefunction factorizes into a set of single-particle wavefunctions, the motions of each particle are independent. The second is that, when the configuration space wavefunction decomposes into a sum of non-overlapping packets, the system point (representing the *N* particles) moves within a domain occupied by just one packet, the other packets being "empty".

For this system, the guidance equation of the *n*th particle, which comes out of the non-relativistic limit of the many-body Dirac equation [12], is given by

$$m_n \dot{\mathbf{x}}_n = (\hbar/2i\rho)(\phi^+ \nabla_n \phi - (\nabla_n \phi^+)\phi) + (1/\rho)\nabla_n \times (\rho \mathbf{s}_n), \quad \rho = \phi^+ \phi, \quad \mathbf{s}_n = (\hbar/2\rho)\phi^+ \sigma_n \phi. \quad (6.1)$$

Here in the expression for the *n*th spin vector $\mathbf{s}_n$ we trace over the spin indices for $n' \neq n$. This law generally implies nonlocal correlations among all the particles. However, when the wavefunction factorizes,

$$\phi^{A_1...A_N}(\mathbf{x}_1,...,\mathbf{x}_N) = \prod_n \phi_n^{A_n}(\mathbf{x}_n), \quad (6.2)$$

the formula (6.1) reduces to the single-particle equation (2.9) for each *n*, which proves our first assertion. Next, suppose the wavefunction decomposes as follows:

$$\phi = \sum_a \phi_a, \quad \phi_a \cap \phi_{a'} = 0, \quad a' \neq a \quad (6.3)$$

Then it is obvious that the right-hand side of (6.1) likewise decomposes and the set of particle velocities $\dot{\mathbf{x}}_n$ are determined by just one of the summands in (6.3). This proves our second assertion and we conclude that the model is empirically sufficient.

## 7 Remarks on the choice of wavefunctions and ensembles

### 7.1 Non-relativistic wavefunctions

In order that the solutions of the Pauli equation we use are limiting forms of Dirac solutions, they must obey conditions corresponding to the conditions assumed in obtaining the Pauli equation from the Dirac equation. A way of gauging whether a wavefunction indeed pertains to the non-relativistic regime is to examine the corresponding particle speed that it implies via the guidance formula, and check that this is small compared to the speed of light. It turns out that the condition on the Dirac wavefunction involved in the method of eliminating small components [21] is actually equivalent (for the wavefunctions we use) to this requirement.

Writing the Dirac spinor as $\psi = \begin{pmatrix} \phi \\ \varphi \end{pmatrix} \exp(-imc^2 t/\hbar)$, Dirac's equation in the low energy limit implies

$$\varphi = \frac{1}{2mc} \boldsymbol{\sigma} \cdot (-i\hbar \nabla - \mathbf{A})\phi \tag{7.1}$$

with

$$|\phi| \gg |\varphi|, \tag{7.2}$$

where $|\phi| = \sqrt{\phi^+ \phi}$ and $|\varphi| = \sqrt{\varphi^+ \varphi}$. Combining (7.1) and (7.2) gives a condition on the Pauli wavefunctions. For the case of vanishing magnetic field and a spin eigenstate, (2.5), we may express the condition (7.2) in the following equivalent but physically more transparent way, using (2.6) and some simple vector algebra:

$$1 \gg \frac{1}{2}\left( \frac{|\mathbf{j}|^2}{c^2 \rho^2} + \left(\frac{1}{mc}\nabla \log \rho \cdot \mathbf{s}\right)^2 \right)^{1/2}. \tag{7.3}$$

This relation essentially states that the particle velocity (2.9) (= $\mathbf{j}/\rho$) is small compared to the speed of light. Our requirement that we are working in a limiting regime of the Dirac theory may be satisfied therefore by choosing Schrödinger wavefunctions and the attendant parameters such that the particle speed remains subluminal. This condition is satisfied in our calculations (see §§8.2 and 11), where in particular we always have $\nabla \rho \cdot \mathbf{s} = 0$.

**7.2 Canonical ensembles**

The solenoidal field $\mathbf{A}$ in (2.10) can in general be written as $\mathbf{A} = \nabla a \times \nabla b$ where the scalar functions $a(\mathbf{x},t)$ (= $-\log\rho$) and $b(\mathbf{x},t)$ (= $\mathbf{s} \cdot \mathbf{x}$) are known as Euler potentials [22]. The field lines are then given by the intersections of the surfaces of constant $a$ and $b$. As $-\log\rho$ is an Euler potential, the surfaces of constant $\rho$ (as well as the planes $\mathbf{s} \cdot \mathbf{x} = \text{constant}$) are involved in the determination of the trajectories. Thus, the invocation of Lorentz covariance has the effect of putting the surfaces of constant density on the same footing as those of constant phase. The trajectories obtained from the integration of the de Broglie-Bohm guidance relation are the orthogonals to the Hamilton-Jacobi surfaces. By contrast, those derived from the extended guidance relation are a vector sum average of the orthogonals of $S = \text{constant}$ and of $\rho = \text{constant}$. Thus, they, as it were, encode for both the motion of particles and the configuration of the $\rho$ field. Here there is another analogy with electromagnetism: individual electromagnetic field lines represent the acceleration of a test particle yet collectively they map out the overall configuration of the field (though there is a key difference in the way they couple with their respective test particles).

A relevant issue then is the choice of trajectory ensembles that will best represent the overall configuration of the density. Initial conditions for classical trajectories are obtained by variation of parameters on a surface of constant $S$. No extra conditions besides dynamical considerations enter the selection of particular ensembles (in geometrical optics these are the ray congruences). Initial conditions for quantum trajectories, by contrast, can be selected to produce ensembles that map faithfully onto the local density but the choice of the domain of the initial conditions is arbitrary. One domain is nevertheless singled out, namely the one composed of the equally spaced points that lie on a surface of constant density. For convenience we shall refer to this as a canonical ensemble.

To facilitate comparison with the original computations of the double-slit (Fig. 1 [3]) we have maintained some of the original features but have modified others. Thus, for instance, the Gaussian wavefunction still forms the basis of the calculations but the original "pseudo" two-dimensional wavefunctions, where the *x*-coordinate was set proportional to the time (see §3), are replaced by genuine two-dimensional ones. This is necessary in order to show fully the action of the irrotational velocity. Finally, we have adopted a developmental approach whereby the novel effects are examined in cases that build up progressively towards the double-slit. The trajectories have been computed by applying a Runge-Kutta numerical integration of the extended guidance formula using the MATHCAD 2001 program.

## 8 The two-dimensional Gaussian wavefunction – a case of quantum inertial motion

### 8.1 The trajectories and the representation of circulation

As in §3, we suppose that the wavefunction factorizes in the three coordinate directions, with the spin vector in the *z*-direction, so that the guidance equation is (3.2). The *z*-component of the particle motion is then independent of the other components. We work in cylindrical polar coordinates $(x = r\cos\theta,\ y = r\sin\theta,\ z)$. Consider the free wavefunction that is a two-dimensional symmetrical Gaussian wavefunction centred about the origin in the *x*- and *y*-coordinates and whose *z*-dependence is for the present unspecified. The normalized density and phase are, respectively,

$$\rho(r,\theta,z,t)(\equiv \rho_1 \rho_2 \rho_3) = \frac{1}{2\pi\sigma^2} e^{-\frac{r^2}{2\sigma^2}} \rho_3(z,t) \tag{8.1}$$

and

$$S(r,\theta,z,t)(\equiv (S_1 + S_2 + S_3)) = -\hbar \tan^{-1}\left(\frac{\hbar t}{2m\sigma_0^2}\right) + \frac{r^2 \hbar^2 t}{8m\sigma_0^2 \sigma^2} + S_3(z,t) \tag{8.2}$$

Here the half-width is given by

$$\sigma(t) = \sigma_0 \left(1 + \gamma^2 t^2\right)^{1/2}, \quad \gamma = \hbar/2m\sigma_0^2. \tag{8.3}$$

The speed of the wavepacket's centre has been set equal to zero and therefore this function will spread but not propagate. The ensuing trajectories can therefore be considered to be viewed from the wavepacket's rest frame. We begin by examining the character of the Lorentz-like force for this wavefunction.
The spin-dependent component (2.10) of the momentum is

$$\mathbf{A}(r,t) = -\frac{\hbar r}{2\sigma^2} \hat{\boldsymbol{\theta}}. \tag{8.4}$$

Inserting (8.4) and $Q'$ in (4.6), the electric and magnetic fields are

$$\mathbf{E} = \frac{\hbar^2 r}{2m\sigma^4}\left(\hat{\mathbf{r}} - \gamma t \hat{\boldsymbol{\theta}}\right) + \frac{\hbar^2}{2m}\frac{\partial}{\partial z}\left(\frac{1}{\sqrt{\rho_3(z)}}\frac{\partial^2 \sqrt{\rho_3(z)}}{\partial z^2}\right)\hat{\mathbf{s}}, \quad \mathbf{B} = -\frac{\hbar}{\sigma^2}\hat{\mathbf{s}}. \tag{8.5}$$

The cross product of $(\nabla S - \mathbf{A})/m$ with $\mathbf{B}$ gives a force that is equal and opposite to the component of $\mathbf{E}$ in (8.5) lying in the *xy*-plane, for all values of *r* and *t*. Thus, the Lorentz force in the *xy*-plane produced by a single two-dimensional symmetrical Gaussian is everywhere zero. Quantum test particles in the Gaussian's null force field must therefore exhibit inertial motion in the plane. If in addition we suppose the *z*-component of $\mathbf{E}$ is zero (this is secured if the wavefunction is plane in the *z*-direction), we obtain inertial motion in three-dimensional space.

This surprising result is easily verified by direct integration of the particle equations of motion (3.2). Separating the radial and angular components of the guidance relation we find for the velocity components

$$\dot{r} = \frac{\gamma^2 rt}{\left(1 + \gamma^2 t^2\right)}, \quad \dot{\theta} = \frac{\gamma}{\left(1 + \gamma^2 t^2\right)}. \tag{8.6}$$

Straightforward integration of (8.6) gives the orbits

$$r = r_0\left(1 + \gamma^2 t^2\right)^{1/2}, \quad \theta = \theta_0 + \tan^{-1}\gamma t \tag{8.7}$$

where $r_0, \theta_0$ are the initial coordinates. From (8.6) and (8.7) it is evident that the magnitude of the speed is a constant for a given trajectory:

$$\left(\dot{r}^2 + r^2\dot{\theta}^2\right)^{1/2} = \gamma r_0. \tag{8.8}$$

The value of the speed is proportional to the trajectory's initial distance $r_0$ from the centre of the wave packet and inversely proportional to the square of the packet's half-width $\sigma_0$. Eliminating *t* from (8.7), the trajectories in the plane are

$$r\cos(\theta - \theta_0) = r_0 \tag{8.9}$$

or, in Cartesian coordinates,

$$y = r_0/\sin\theta_0 - x/\tan\theta_0 \text{ or } yy_0 + xx_0 = y_0^2 + x_0^2. \tag{8.10}$$

These are the equations of linear trajectories whose tangent vectors remain normal to the initial radius vector $\mathbf{r}_0 = r_0\hat{\mathbf{r}}$ for all *t* (from (8.6), the initial velocity components are $\dot{r}_0 = 0, \ \dot{\theta}_0 = \gamma$). Fig. 2 shows an ensemble of trajectories corresponding to initial points located on a circular contour of constant density whose radius is $r_0 = \sigma_0 (= 2 \times 10^{-8} \text{ m})$

To gain further insight into the particle behaviour, we write the vector (8.4) in the following suggestive form

$$\mathbf{A}(r,t) = -m\boldsymbol{\omega}\times\mathbf{r}, \quad \boldsymbol{\omega} = \frac{\hbar}{2m\sigma^2}\hat{\mathbf{s}} \tag{8.11}$$

where **ω** is an angular velocity in the direction of the spin. It is this way of expressing the spin-dependent contribution to the momentum that lies behind Hestenes' [9,10] suggestion that the spin may be attributed to an overall circulatory orbital motion of an electron rather than to an intrinsic property (see §3). However, this circulation is an underlying property of the field and it is not manifested in the motion of an individual particle (see below). Rather, as is clear from Fig. 2, and as implied by (3.4), the circulatory character of the field is represented in the ensemble of trajectories, each of which is inertial but has a finite angular velocity. The overall effect is analogous to the ejection of particles from a (horizontal) Catherine Wheel.

In connection with this, Hestenes, as well as Bohm and Hiley [12], were mistakenly led to expect a spiralling of trajectories in the three-dimensional Gaussian ball case. In fact, the projection of the trajectories there onto a plane orthogonal to the spin direction (the $z$-axis) will remain linear. This is best seen by noting that the equations of motion in spherical polars are independent of the azimuthal angle $\phi$ while the equations in $r$ and $\theta$ are essentially the same as in the two-dimensional case. The solutions therefore are constrained to lie on the intersection of the cone $\phi$ = constant with the family of planes parallel to the $z$-axis. The $z$-component of the orbit is given by $z = z_0 \left(1 + \gamma^2 t^2\right)^{1/2}$ and these hyperbolic trajectories for the Gaussian ball do not exhibit any spiralling.

Although the circulatory character of the field (8.11) is not manifest in the motion of an individual particle, it is nevertheless relevant to an orbit in the way that the orthogonal motions are interdependent. In Cartesian coordinates the orbit (8.7) is

$$x = x_0 - y_0 \gamma t, \quad y = y_0 + x_0 \gamma t. \tag{8.12}$$

We see that the speed in each direction depends on the initial coordinate in the orthogonal direction. This feature of mutual dependence also illustrates that, even though we obtain for this wavefunction the (inertial) motion we would expect in the classical domain (for zero external forces), for finite times this is not an example of the emergence of the classical limit described in §5 (for the quantum potentials $Q'$ and **A**, and the **E** and **B** fields, are finite). On the other hand, as $t \to \infty$ it is easy to see from (8.1)-(8.5) that $\nabla S$ tends to a constant, and $Q'$, **A**, **E** and **B** go to zero. Thus, in this limit the conditions for the classical limit are in fact obeyed (and the spin $\tfrac{1}{2}$ orbits tend towards the de Broglie-Bohm spin 0 ones (see below)).

It is useful to note the mechanism by which the spin-dependent term alters the trajectories determined by the de Broglie-Bohm guidance equation (1.1) while leaving the density unaffected. For this wavefunction, the de Broglie-Bohm orbits are also linear in space, but radial (the angular component of the velocity is zero):

$$xy_0 = yx_0. \tag{8.13}$$

Along the tracks the speed is variable. Let us choose again the initial points to lie on a circular contour of constant density. At a given time $t$, a two-dimensional infinitesimal element $rdrd\theta$, fixed in space, will be traversed by trajectories that intersect orthogonally the arcs of the element. Thus the net flux occurs only over the curved sides and arises out of the $r$ dependence of the velocity. "Switching on" the **A**-component of the velocity produces trajectories that will traverse the same element across all four sides. These trajectories will start from different initial locations than the de Broglie-Bohm ones, but on the same contour of constant $\rho$. However, because the angular component of the velocity (8.6) is independent of $\theta$ the net flux contribution will be as before and therefore the rate of change of density is unaltered by the addition of the solenoidal field.

In conclusion, we observe that in the restricted dynamics of de Broglie and Bohm the role of generator of inertial motion is fulfilled by the three-dimensional plane wave. In the extended theory the plane wave still implies inertial motion, but we have the new feature that such motion is also generated by a two-dimensional Gaussian wavefunction that is a plane wave in the direction of the spin vector. The inertial property is a consequence of the kinetic interdependence of the orthogonal degrees of freedom described in §3. The latter is such that the acceleration implied by the original de Broglie-Bohm guidance equation for this wavefunction is cancelled.

**8.2 Remarks on the non-relativistic approximation**

The proportionality between the speed and the initial radial distance from the centre of the packet raises the question of the extent to which our model may be regarded as representing a genuine limiting case of the relativistic theory, for which motions will remain subluminal (see §7.1). This problem may be posed in probabilistic terms: what is the likelihood that there exist trajectories that are initially located so far away from the centre of the packet as to make their speed approach or exceed $c$. Using the normalized density (8.1) (ignoring the $z$-component), the fraction of initial points lying between $r_0$ and $r_0 + \delta r_0$ is $2\pi \rho_0(r_0) r_0 \delta r_0$. From (8.8) this can be converted into the fraction of trajectories whose speed lies between $v$ and $v + \delta v$:

$$\frac{\delta n}{n} = \frac{v}{w^2} e^{-\frac{v^2}{2w^2}} \delta v \tag{8.14}$$

where $n$ is the total number of trajectories and $w = \gamma \sigma_0 = \hbar/2m\sigma_0$ is a characteristic speed associated with a particular Gaussian. The distribution (8.14) is peaked about $v = w$ and we see that, although the speed increases linearly with $r_0$, the fraction of trajectories corresponding to speeds above the characteristic value is a rapidly decreasing function. Thus, choosing $w$ suitably, the restriction on speeds approaching luminal or superluminal values results from the vanishingly small likelihood that they will be physically realized. This is indeed the case for the calculations that we present in this paper where the characteristic speed has the value 3.7 x $10^4$ m/s. From (8.8) it follows that we must choose only those initial conditions for which $r_0 < c/\gamma \approx 10^{-4}$ m. Hence, for the bulk of the trajectories (and certainly for the ones we consider here) our model is a good non-relativistic approximation.

**9 Trajectories in the laboratory frame**

Having considered the trajectories in the rest frame of the Gaussian wave packet, we next examine their appearance in a laboratory frame when the centre of the wave packet is propagating at a constant velocity **u**. This amounts to finding the Galilean boosted trajectories. In general, under the transformation $\Psi(\mathbf{x}) \to \Psi'(\mathbf{x} + \mathbf{u}t)$ the trajectories transform as [20]

$$\mathbf{x}'(t) = \mathbf{x}(t) + \mathbf{u}t. \tag{9.1}$$

The ensemble of Gaussian trajectories, labelled by different initial position vectors $\mathbf{r}_0 = (x_0, y_0)$ relative to the centre of the packet, are, from (8.12), given by

$$\mathbf{r}(t) = \mathbf{r}_0 + \mathbf{v}t \tag{9.2}$$

where $\mathbf{v} = \gamma r_0 \hat{\mathbf{b}}$ and $\hat{\mathbf{b}} \, (\propto (-y_0, x_0))$ is a unit tangent vector normal to $\mathbf{r}_0$. From (9.1) the new orbits will therefore be given by

$$\mathbf{r}'(t) = \mathbf{r}_0 + (\mathbf{v} + \mathbf{u})t. \tag{9.3}$$

The trajectories remain linear but with constant unit tangent vectors given by

$$\hat{\mathbf{b}}' = \frac{\gamma r_0 \hat{\mathbf{b}} + \mathbf{u}}{\sqrt{\gamma^2 r_0^2 + \mathbf{u}^2}}. \tag{9.4}$$

Thus, in the laboratory frame, the straight trajectories appear rotated through angles $\alpha$ that are a function of $r_0$ and $\beta$, the latter being the angle between the trajectory in its rest frame and the direction of motion of the centre of the wave packet:

$$\alpha(r_0, \beta) = \cos^{-1}\left(\frac{\gamma r_0 + u\cos\beta}{\sqrt{\gamma^2 r_0^2 + \mathbf{u}^2}}\right). \tag{9.5}$$

It can be seen that in the limit $u \gg \gamma r_0$ of high packet speed, $\alpha \to \beta$. In this limit the trajectories are rotated by an amount equal to the angle they make initially with the direction of motion of the wave packet. For instance, if the wave packet is moving in the *x*-direction, the trajectories starting off at $\theta = 0$, $\pi/2$, $\pi$ and $3\pi/2$ will appear, in the laboratory frame, to be rotated through $\alpha = \pi/2$, $\pi$, $3\pi/2$ and $2\pi$, respectively.

Fig. 3 shows the same ensemble of trajectories as in Fig. 2 as seen in the laboratory frame for a packet motion $\mathbf{u} = (V_x, 0)$, for three different values of $\gamma r_0/u$. The constant packet speed $V_x$ does not destroy the inertial motion but it does break the symmetry between the trajectories that travel in the same direction and those that are travelling opposite to it. The trajectories are seen to cross as a result of this, but this is not a violation of the single-valuedness of the wavefunction since they reach the same points in space at different times. A symmetrical distribution of the ensemble is reached asymptotically as $V_x$ becomes much greater than the trajectory speed. These features will be encountered again in the double-slit.

## 10 Products and superpositions of Gaussian packets

We now pass to consideration of some of the key differences between the motions obtained when the wavefunction is a product or a superposition, in situations where the Lorentz force is finite. The principal difference between the two cases is the emergence of interference effects in the case of superposition. We shall ignore the *z*-component of the motion.

We start with the case when $\Psi$ is a product of two stationary Gaussian functions of *x* and *y*, of different initial half-widths $\sigma_{0x}$ and $\sigma_{0y}$, respectively. The contours $\rho$ = constant will now be elliptical and a canonical ensemble will be obtained from initial conditions that are evenly distributed on an ellipse whose semi-major axes are $\sigma_{0x}$ and $\sigma_{0y}$. Calculation of the electric-like force shows that it is not equal and opposite to the magnetic-like force along the trajectory as was the case with the symmetric Gaussian. Consequently, the Lorentz-like force no longer vanishes and the resulting trajectories exhibit acceleration. Their departure from linearity depends on the eccentricity $e = \sqrt{\sigma_{0x}^2 - \sigma_{0y}^2}/\sigma_{0x}$ of the initial canonical ensemble. In this instance, the kinetic interdependence expresses the anisotropy of the asymmetrical product wavefunction.

Fig. 4 shows the trajectories in their early stage of motion and as they develop over a longer period of time, for the case where $\sigma_{0x} = 2\sigma_{0y}$. Deviations from linearity, and hence accelerations, occur early on but as time progresses the paths get ironed out into straight lines, indicating that the Lorentz-like force decays with time. In fact, as in §8.1, in this limit the orbits obey the classical equations.

We shall now examine the canonical ensembles of trajectories arising from the superposition wavefunction

$$\Psi(x,y,t) = N\eta(x,t)[\xi(y-a,t) + \xi(y+a,t)] \tag{10.1}$$

where $\eta$ and $\xi$ are stationary Gaussian functions of their arguments with the same half-widths and $N$ is a normalization constant. Clearly, when the separation $2a$ between the two component wavefunctions $\xi$ is zero, $\Psi(x,y,t)$ reduces to a single symmetric Gaussian. The Lorentz force therefore will vanish everywhere and the trajectories will be as in Fig. 2. Imagine next an infinitesimal increase in the separation between the two Gaussians of the superposition. The total wave packet will undergo an asymmetrical deformation so that it elongates in the $y$-direction. In a first approximation, the contours of constant $\rho(=|\Psi|^2)$ may be considered to be ellipses centred on the origin. The situation will be similar to the one considered above for the product state and the Lorentz force will become a finite function of position. The resulting trajectories will deviate from linearity and a canonical ensemble will be similar to those shown in Fig. 4.

Further increase in the value of $a$ will result in a decrease of overlap and a gradual resolution of the wavefunction into two separate peaks. The contours of constant $\rho$ will show the new topological feature of sets of pairs of closed curves centred on each peak at $(0, \pm a)$. These curves tend towards circles as the overlap between the Gaussians is made to diminish. The shape of the trajectories associated with these pairs of contours is best understood by examining the symmetry properties associated with the convective and circulatory velocities. To this end, Fig. 5 compares the trajectories when the spin-dependent vector **A** is turned off and then switched on (the plots are for separation $2a = 5\sigma_0$ and initial ensemble lying on the contours at $0.5\sigma_0$).

The functional dependence of $(\partial S/\partial x, \partial S/\partial y)$ on $x$ and $y$ is of degree one and the system of trajectories is symmetrical with respect to reflections in the $y$- and $x$-axes. Adding the circulation introduces a chirality that alters this symmetry. Since the spin is in the $z$-direction, the **A**-field in the vicinity of each peak approximates to an anticlockwise rotation centred on each peak. Its effect therefore is to "sweep" each set of trajectories in the same sense. Fig. 5 shows how the set originating from the upper wave packet is swept towards the left by the rotation centred on the lower wave packet, and vice versa. The resulting asymmetry about the $x$-axis is of key importance in what follows.

The deviation from linearity now is of a fundamentally different kind from that of the product wavefunction shown in Fig. 4 but it is still accounted for by the Lorentz-like force. Both the electric-like and magnetic-like components of the force are functions of the phase difference of $\xi(y-a,t)$ and $\xi(y+a,t)$ and therefore the trajectories of the superposition show the features that are indicative of interference. The rapid deviations over short distances, or kinks, seen in the first picture of Fig. 5 occur in regions where the overlap between the wave packets is sufficient to make the phase difference significant. These develop into loops, visible in the second picture, when the circulation is added. In Fig. 5 the wave packets were chosen to be initially close to each other in comparison with their half-width, and the interference kinks appear relatively early. The trajectories become linear asymptotically as the spreading wave packets merge into each other in time to simulate a single Gaussian. Conversely, packets that are initially far apart have trajectories that start off being linear and in time, as the overlap begins to come into effect, they develop the characteristic interference loops.

# 11 The two-slit experiment

We shall end by showing how the trajectory interpretation of the two-slit interference experiment is modified by the inclusion of the vector potential in the guidance formula. The double-slit is modelled by two identical two-dimensional symmetrical Gaussian packets travelling in the *x*-direction at a speed $V_x = 10^8$ m/s. This value is chosen so as to ensure that the trajectories reach the Fraunhofer limit within the time of the calculation. At $t = 0$ the packets have half-widths $\sigma_0 = 2 \times 10^{-8}$ m and their separation is $20\sigma_0$. Thus the contours of constant $\rho$ can be approximated by circles up to distances of the order of $\sigma_0$. The canonical ensemble consists of points located on six concentric circles whose radii are multiples of $0.4\sigma_0$. The total number of points in the ensemble is normalized so as to give 20 points at $\sigma_0$ while the number of points in each subensemble is the nearest integer calculated from the value of $\rho$ on the corresponding contour.

Figs. 6 and 7 show the trajectories obtained by numerically integrating the new guidance relation with $\mathbf{A}(\mathbf{x}.t) = 0$ and $\mathbf{A}(\mathbf{x}.t) \neq 0$, respectively. In both cases the trajectories exhibit the same bunching effect in the far-field limit as those of the original calculation (Fig. 1 [3]). However, the trajectories in Fig. 7 differ in two key respects: they cross each other, and they also cross the *x*-axis. It will be recalled from the discussion in §9 that intersecting trajectories meet at different particle arrival times and do not pose any difficulties with regards to single-valuedness. For the same reason, though it is not apparent on the scale of the plot, the end-points in the far field at a given time are distributed over the *xy*-plane, and are not confined to a line *x* = constant.

By contrast, the crossing of the axis of symmetry is a more significant novel feature associated with the vector potential. It occurs because of the combined effects of the rotation due to the Galilean boost (§9) and the reciprocal "sweep" (§10) of the separate packets, as illustrated in Figs. 3 and 5. It will be recalled that the magnitude of these effects depends, among other things, on the trajectory speed $\dot{\mathbf{x}} = \gamma r_0$ (which is correlated with the initial position), on the wavefunction speed $V_x$, and on the separation between the packets. Thus the initial location of the trajectory controls whether or not it will cross the axis of symmetry. Clearly, then, the relative localization of particles emanating from each slit in Figs. 1 and 6, in which they are confined to their respective half-planes, is removed by the action of the vector potential. But it needs to be stressed once more that such changes in the trajectory configuration do not affect the particular $|\Psi|^2$ density distribution; they have a bearing only on how we read the process that causes that particular distribution.

Finally, two notes concerning these results. First, the asymmetry in the distribution that can be seen with a few of the trajectories in Fig. 7 is an artefact. It occurs because some of the initial points are asymmetrically distributed on a contour, which happens whenever the nearest integer calculated from the value of $\rho$ is odd. This artefact will of course disappear in the theoretical limit of a continuously distributed ensemble. Second, we have already indicated in §8.2 that the speeds generated by a single Gaussian do not exceed the limits of validity of the non-relativistic calculation (i.e., they remain subluminal). In order to attain the Fraunhofer limit, the group velocity $V_x$ of the two wave packets in the superposition has been chosen to be much greater than the trajectory speeds. Therefore, a final check is in order to ensure that the onset of interference does not result in superluminal speeds. We have computed the speed along a trajectory for a number of trajectories originating on a contour at $1.5\sigma_0$. Fig. 8 shows the variation of the ratio of this speed to the group velocity along each of the trajectories. The variations of the speeds due to the interference effects are clearly visible but they are negligible in comparison with the arbitrarily chosen group velocity. Clearly, the forces that produce the kinks in the trajectories are large but the time over which they act is short.

## 12 Conclusion

Given the emphasis placed on ontology by the de Broglie-Bohm theory it is natural to raise the question whether the current as given by the original non-relativistic guidance formula represents an actual flow of matter. This is a problem since the original theory is underdetermined due to a gauge freedom in the current. On the other hand, for the relativistic spin $\frac{1}{2}$ case the Dirac current does not exhibit a gauge uncertainty and by invoking its low energy limit we obtain a physical principle which fixes the current in the non-relativistic domain. This then gives the current an ontological status it would have otherwise lacked, and the modified guidance relation thus obtained is therefore to be understood as a completion of the de Broglie-Bohm theory. The consequences of this completion for the trajectories corresponding to Gaussian wavefunctions, culminating with those of the double-slit, were examined.

A key feature of the approach is that it places the phase and amplitude of the wavefunction on an equal dynamical footing, a modification that produces some significant changes to the reading of the de Broglie-Bohm theory. First, it revises the trajectory representation of two-slit interference by showing that the trajectories from each slit do in fact cross and that they do not necessarily remain confined to their respective half-planes. More generally, it reveals the presence of a coupling associated with product wavefunctions. Superposition in quantum mechanics generally results in mutual dependence of degrees of freedom that classically would be expected to be statistically independent. The revised guidance relation exhibits mutual dependence of degrees of freedom that quantum mechanically are expected to be statistically independent. This was referred to as kinetic interdependence and is essentially the reason behind the crossing of the trajectories for a superposition. The presence of the amplitude in the guidance equation also implies that, while the translational classical limit of the spin $\frac{1}{2}$ theory coincides with the classical limit of the spin 0 theory, the former involves conditions additional to those involved in the latter. Lastly, the approach introduces some basic and suggestive analogies with electrodynamics. These extend the visualisations and the modes of description of the de Broglie-Bohm theory, in particular by bringing to the fore the notion of electric-like and magnetic-like lines of force, analogous to Faraday's original representation. Whether they carry a deeper theoretical and practical significance will be the subject of future work.

**Figure Captions**

1. Two-slit trajectories calculated using the de Broglie-Bohm law (1.1) [3]. The wavefunction is a superposition of Gaussian packets in the *y*-direction, centred on B and B', and a plane wave in the *x*-direction.

2. The quantum Catherine Wheel. Trajectories for a symmetric two-dimensional Gaussian packet with $r_0 = \sigma_0$.

3. Single Gaussian packet moving in the *x*-direction at speeds (from the top) 0.8, 2 and 5 times the trajectory speed, respectively (same scale as Fig. 2).

4. Trajectories for an asymmetric two-dimensional Gaussian packet with $\sigma_{0x} = 2\sigma_{0y}$ near $t = 0$ and as $t \rightarrow \infty$.

5. Trajectories for the superposition of two two-dimensional Gaussian packets with separation $2a = 5\sigma_0$ and initial ensemble on contours at $0.5\sigma_0$, before and after switching on the vector potential **A**.

6. Two-slit trajectories without the vector potential **A**. The wavefunction is a superposition of two two-dimensional Gaussian packets each of group speed $V_x$. The ensemble at each slit consists of six equally spaced concentric rings whose radii range from $0.4\sigma_0$ to $2.4\sigma_0$. The number of initial points on each ring is weighted by the value of $\rho$ and normalized to 20 at $\sigma_0$.

7. Two-slit trajectories with the vector potential **A**. The ensemble is as in Fig. 6.

8. Ratio of trajectory speeds to wave packet speed $V_x$ for a set of 15 trajectories from one slit. The sub-ensemble belongs to the constant $\rho$ contour at $\sigma_0$.

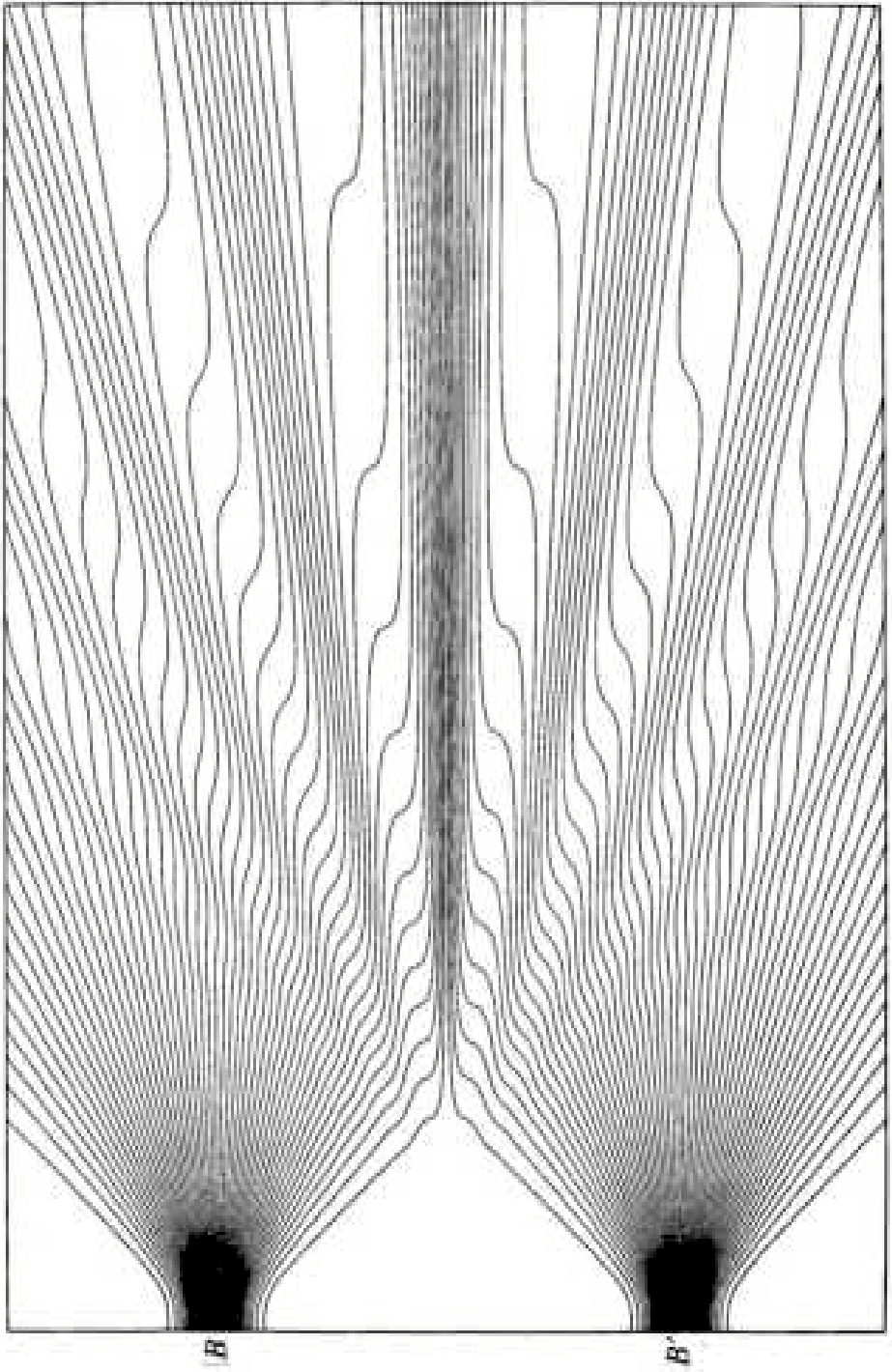

Figure 1

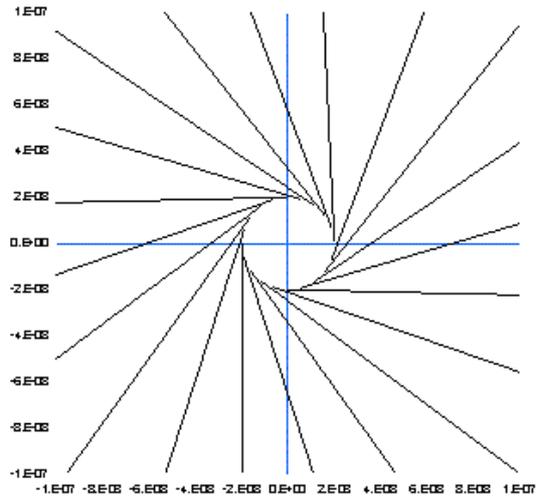

Figure 2

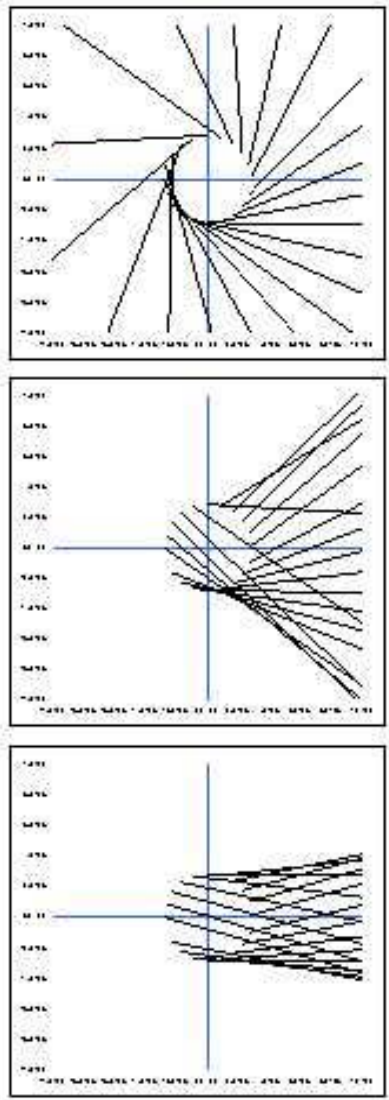

Figure 3

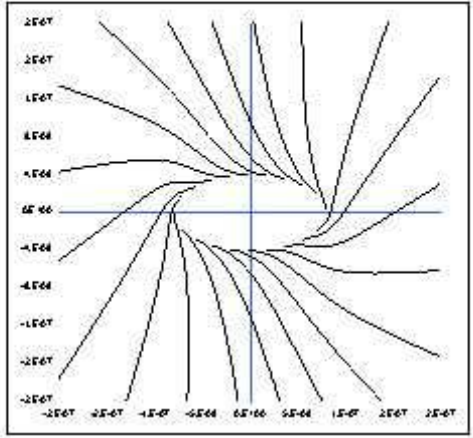
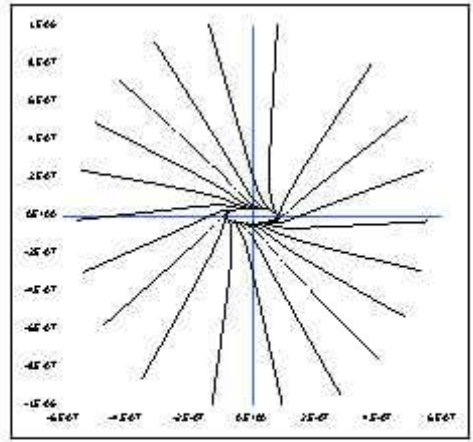

Figure 4

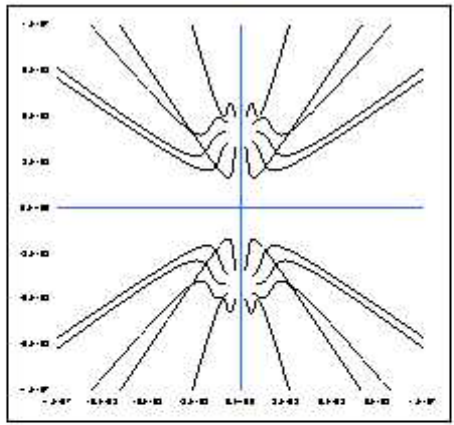
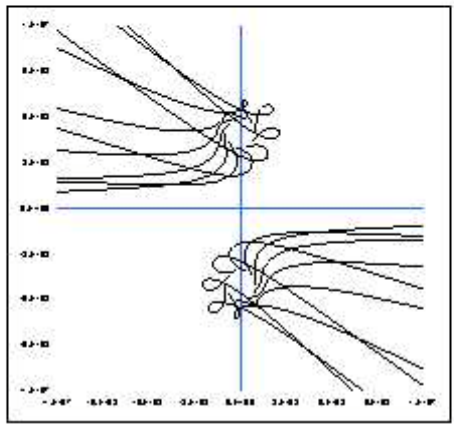

Figure 5

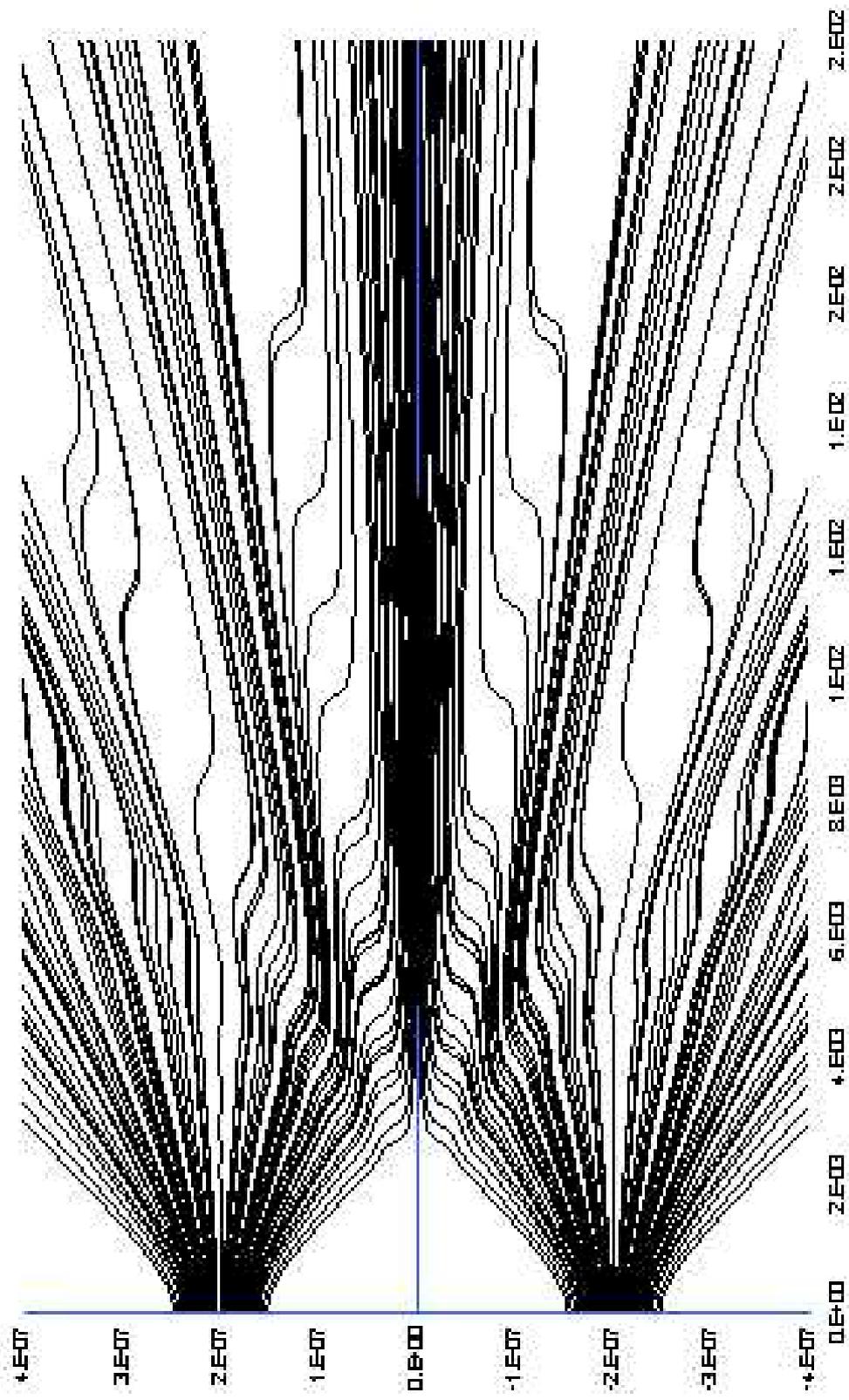

Figure 1

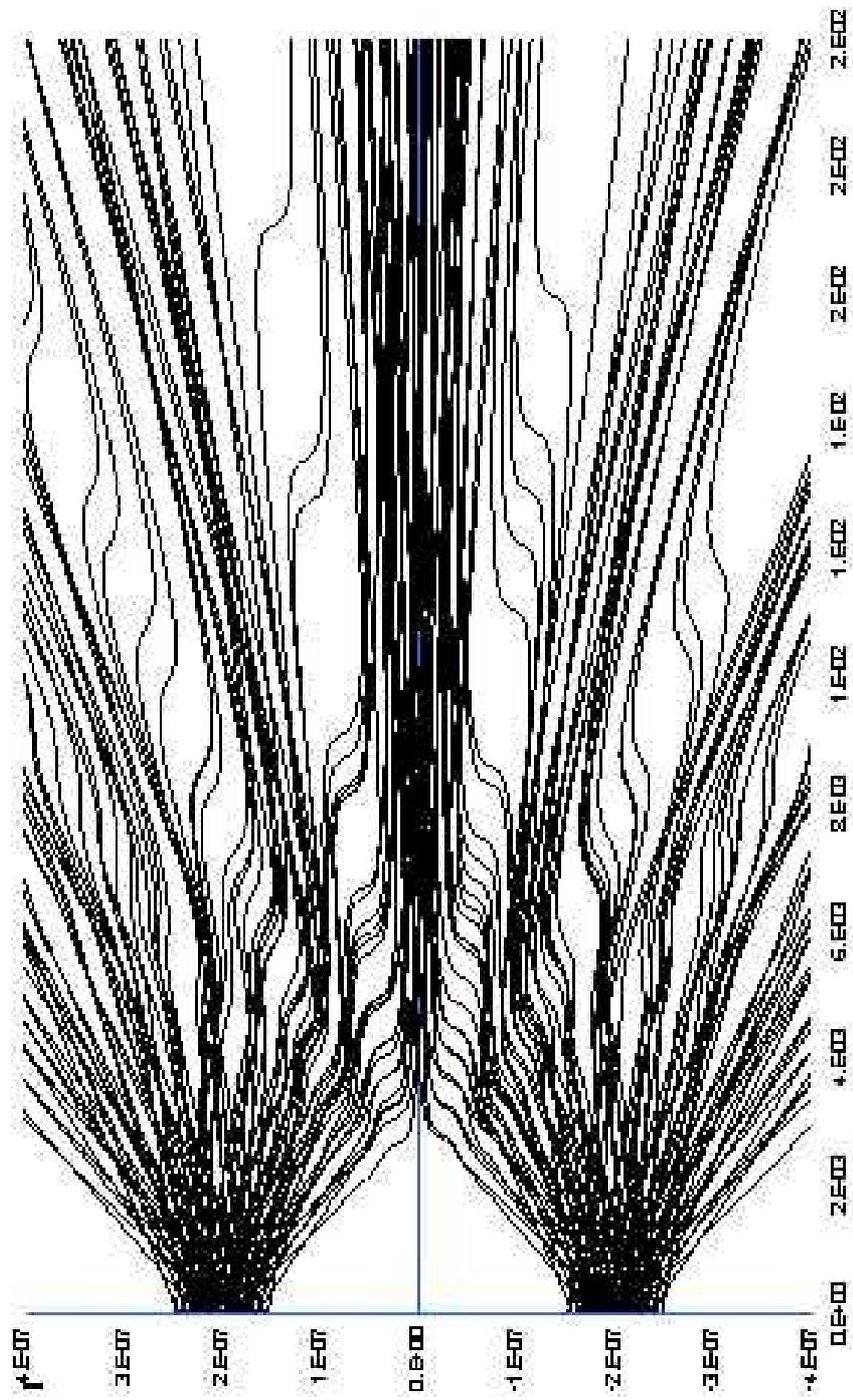

Figure 7

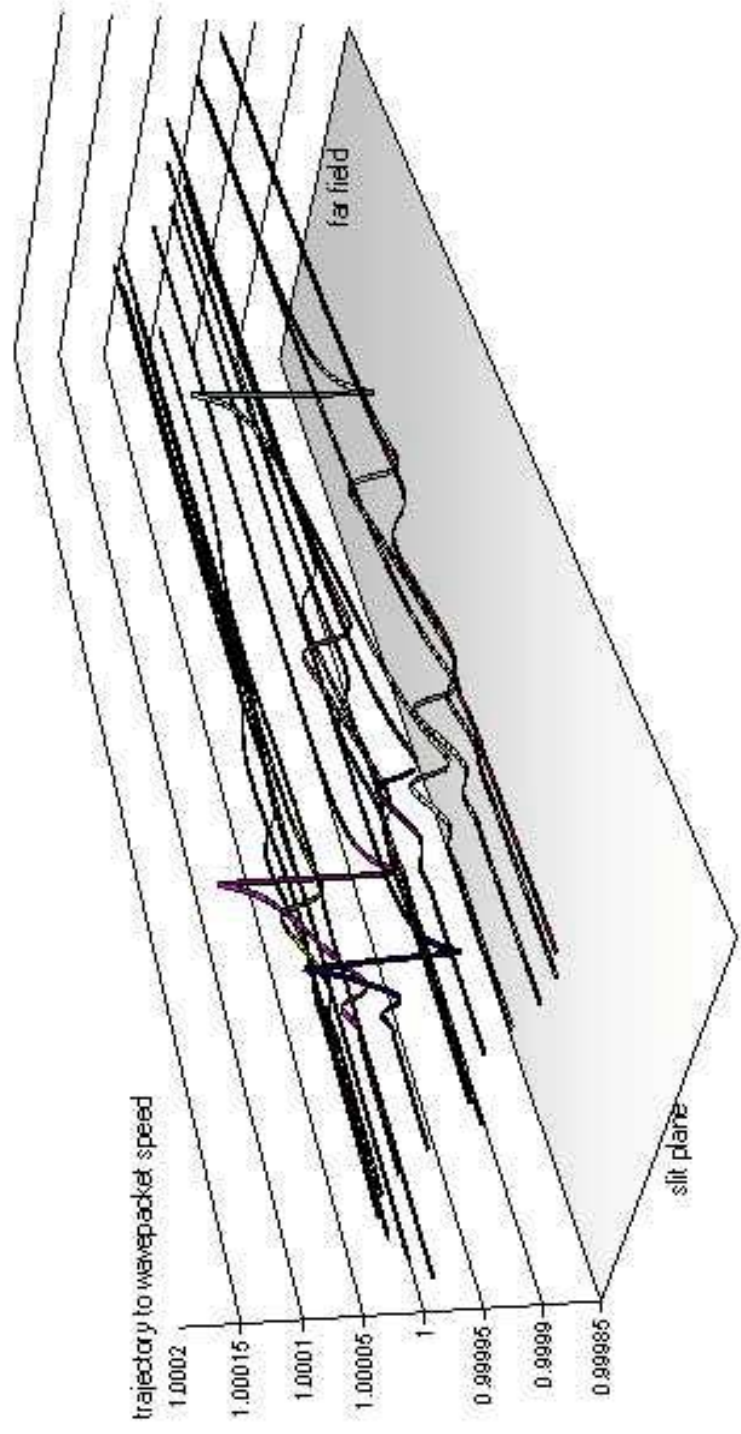

Figure 8